\documentclass{ws-mpla}
\newcommand{\oh}{{1\over2}}
\begin{document}

\markboth{Authors' Names}
{Instructions for Typing Manuscripts (Paper's Title)}

\catchline{}{}{}{}{}

\title{RECENT DEVELOPMENTS IN CHIRAL UNITARY
DYNAMICS OF RESONANCES
}

\author{\footnotesize E. Oset, L. S. Geng, D. Gamermann, M.J. Vicente Vacas, D.
Strottman, K. P. Khemchandani and A. Martinez Torres\footnote{ oset@ific.uv.es}}

\address{Departamento de F\' isica Te\' orica and IFIC, 
Centro Mixto Universidad de Valencia-CSIC, 
Institutos de Investigaci\'on de Paterna, Aptd. 22085, 46071 Valencia, Spain.\\
}

\author{J. A. Oller and L. Roca}

\address{Departamento de F\'{\i}sica. Universidad de
Murcia. E-30071 Murcia.  Spain.\\
}

\maketitle

\pub{Received (Day Month Year)}{Revised (Day Month Year)}

\begin{abstract}
In this talk I summarize recent findings made on the description of axial vector
mesons as dynamically generated states from the interaction of peseudoscalar
mesons and vector mesons, dedicating some attention to the two $K_1(1270)$
states. Then I review the generation of open and hidden charm scalar and axial 
states. Finally, I present recent results showing that the low lying $1/2^+$
baryon resonances for S=$-1$ can be obtained as bound states or resonances of two mesons
and one baryon in coupled channels dynamics.

\end{abstract}

\ccode{PACS Nos.: include PACS Nos.}

\section{Introduction}
	
The combination of nonperturbative unitary techniques in coupled channels with
the QCD information contained in the chiral Lagrangians has allowed one to extend
the application domain of traditional Chiral Perturbation theory to a much
larger range of energies where many low lying meson and baryon resonances
appear. For instance, for the interactions between the members of the lightest 
octet of pseudoscalars, one starts with the chiral Lagrangian of 
ref. \cite{gasser,ulf} and selects the set of channels that couple to certain 
quantum numbers. Then, independently of using either the Bethe-Salpeter 
equation in coupled channels \cite{npa}, the N/D method \cite{nsd} or the 
Inverse Amplitude one \cite{ramonet}, 
the well known scalar resonances $\sigma(600)$, $f_0(980)$, $a_0(980)$ 
and $\kappa(800)$ appear as poles in the obtained L=0 meson-meson partial 
waves.  These
resonances are not introduced by hand, they appear naturally as a consequence of
the meson interaction and they qualify as ordinary bound states or resonances
in coupled channels. These are states that we call dynamically generated, by
contrast to other states which would rather qualify as $q \bar{q}$ states, such as
the $\rho$.  Similarly, in the baryon sector, the interaction of the pseudoscalar
mesons with baryons of the octet of the proton generates dynamically $1/2^-$
resonances \cite{weise,angels,ollerulf,carmina,jido,ollersolo} and the interaction of the
pseudoscalar mesons with baryons of the decuplet of the $\Delta$ generates $3/2^-$
resonances \cite{lutz,sarkar}. These last two cases can be unified using SU(6)
symmetry as done in \cite{carmenjuan}. This field has proved quite productive
and has been further expanded by combining pseudoscalar mesons with vector mesons,
which lead to the dynamical generation of axial vector mesons like the
$a_1(1260)$, $b_1(1235)$, etc \cite{lutzaxial,roca}. Also in the charm sector
one has obtained in this way scalar mesons with charm, like the $D_{s0}(2317)$
\cite{koloscalar,hofmann,chiang,daniel}, axial vector mesons with charm like the
$D{s1}(2460)$ \cite{lutzaxial,chiangaxial,danielaxial}, or hidden charm scalars
like a predicted $X(3700)$ state and two hidden charm axial states, with
opposite C-parity, one of which corresponds to the $X(3872)$ state. In what
follows we briefly discuss these latter cases. 

\section{Axial vector mesons dynamically generated}

\begin{figure*}[ht]
\begin{center}
\begin{tabular}{cc}
\includegraphics[scale=0.25]{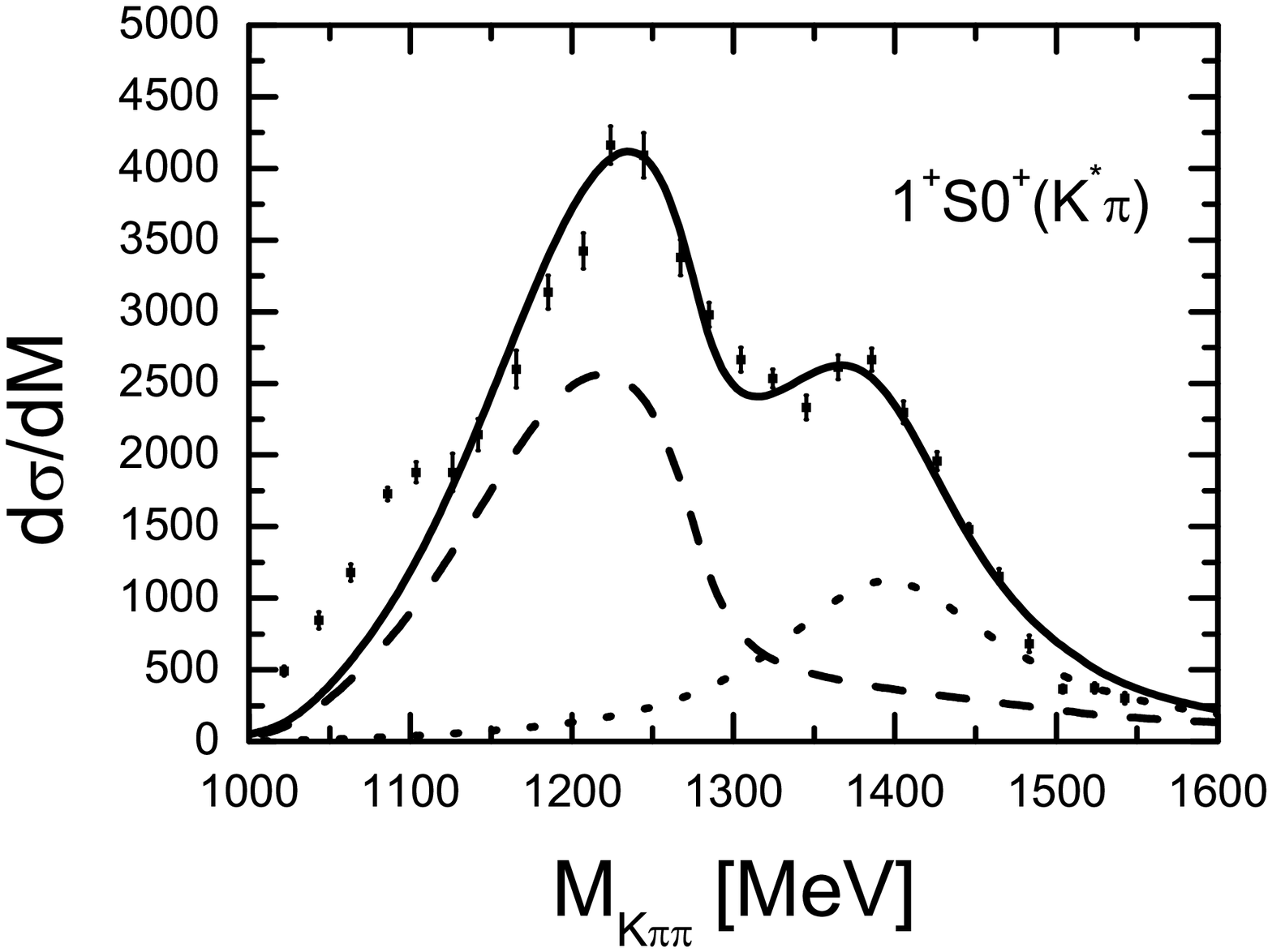} %
&\includegraphics[scale=0.25]{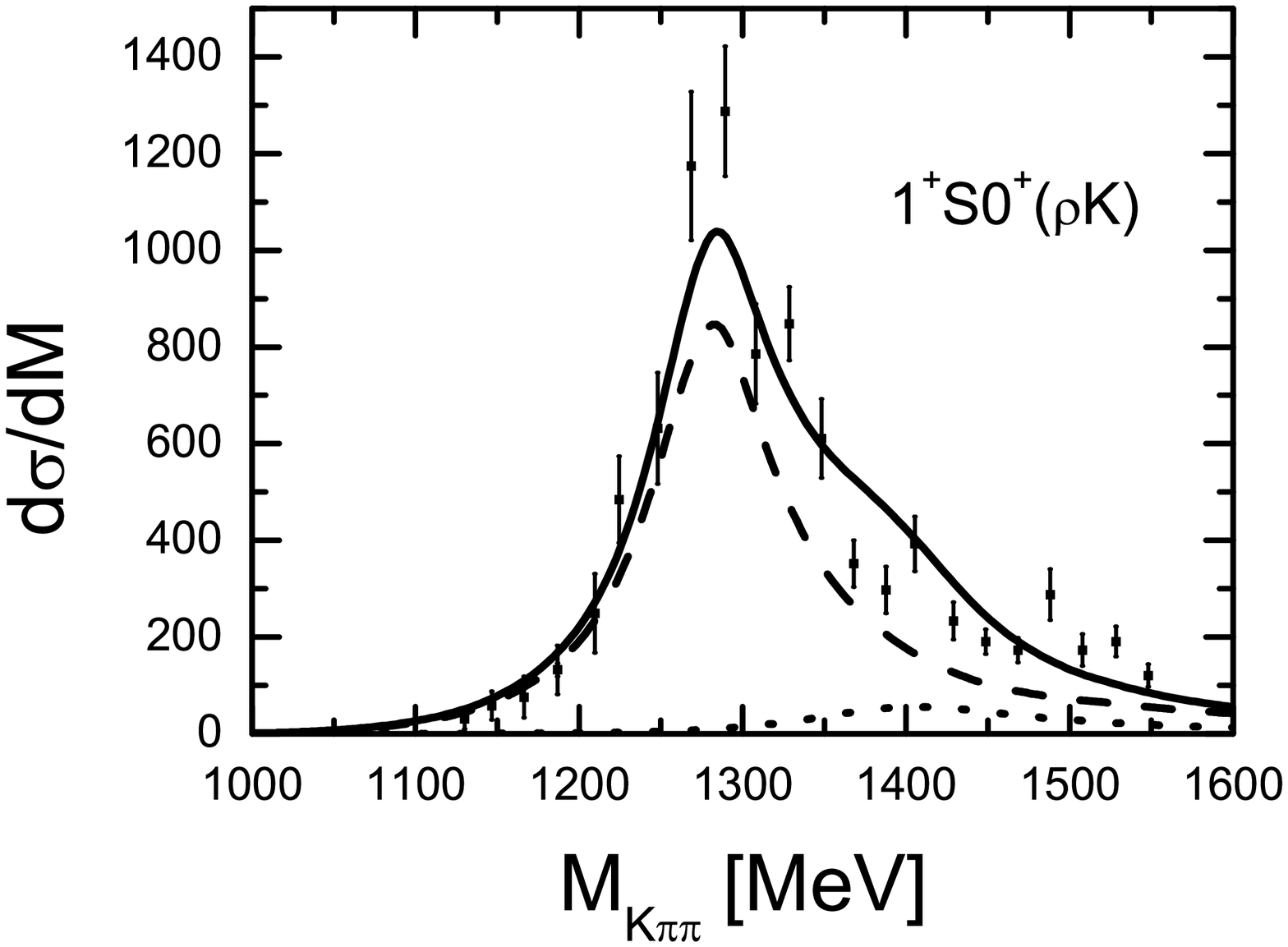}\\
\includegraphics[scale=0.25]{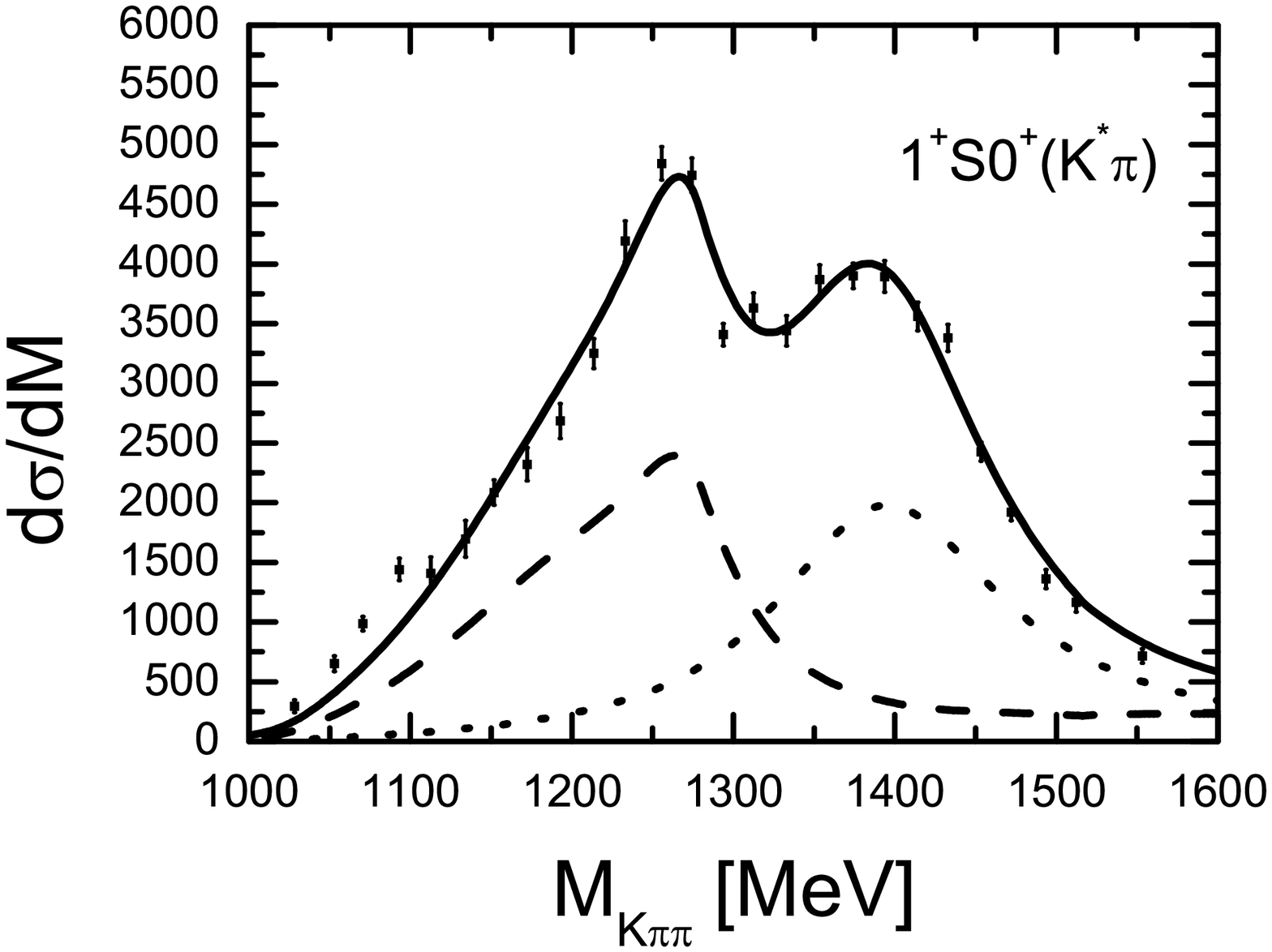}%
&\includegraphics[scale=0.25]{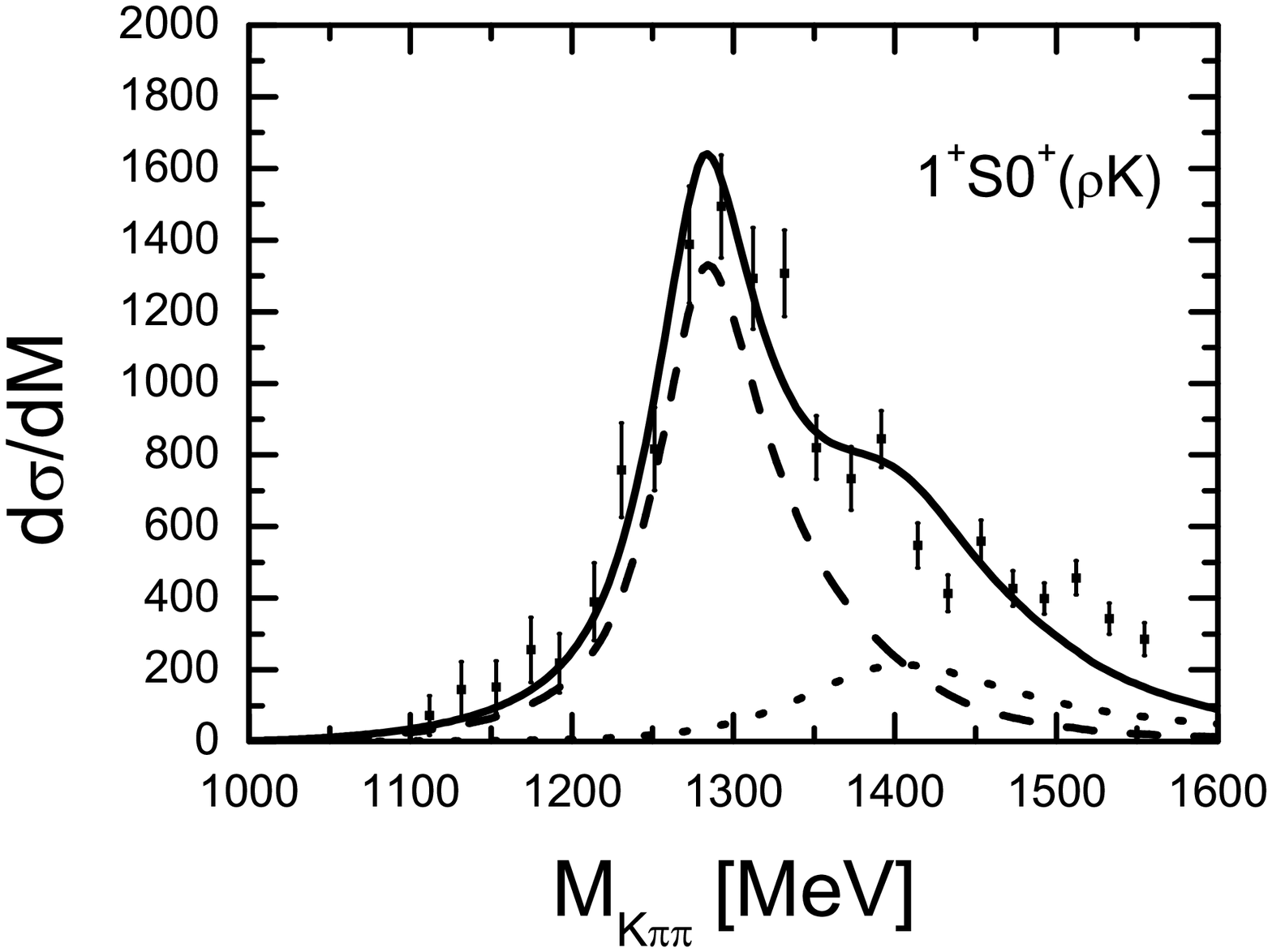}\\
\includegraphics[scale=0.25]{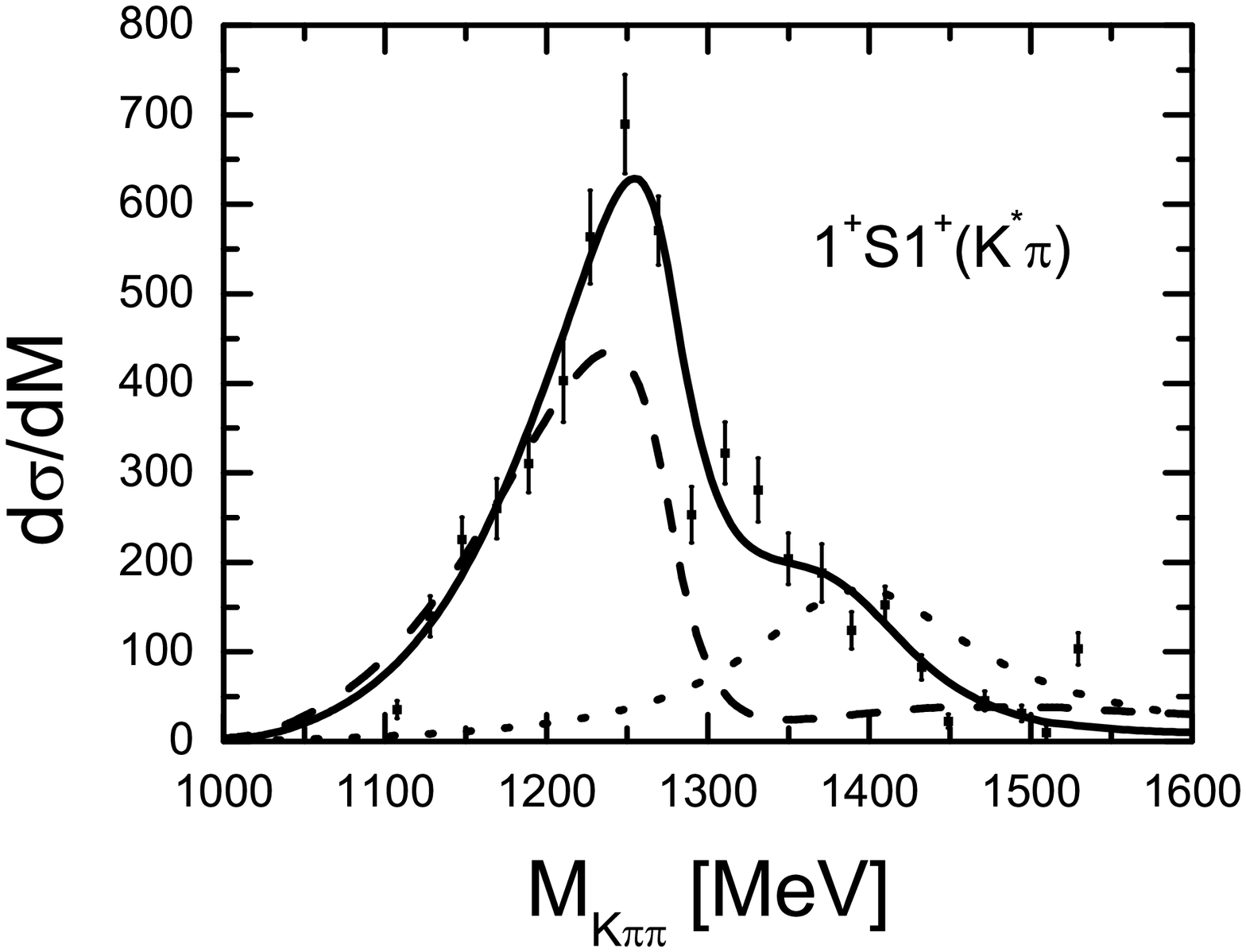}%
&\includegraphics[scale=0.25]{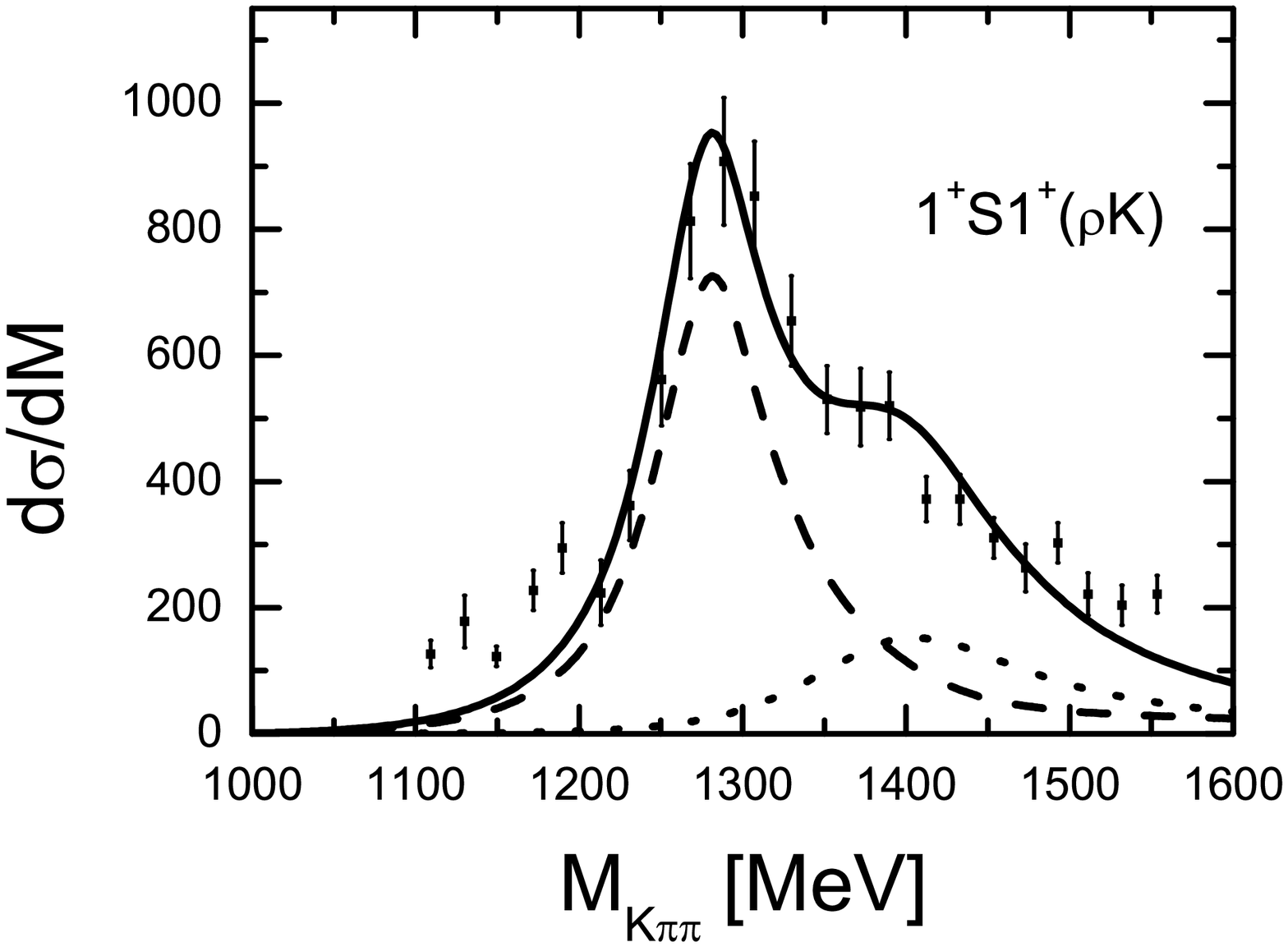}\\
\end{tabular}
\caption{Results for the $ \pi \pi K$ invariant mass distribution in the 
$K^- p \to K^- 	\pi^+  \pi^- p$  reaction. }
\label{fig:CERN3}
\end{center}
\end{figure*}

As shown in detail in \cite{lutzaxial,roca}, starting from  a standard chiral
Lagrangian for the interaction of pseudoscalar mesons of the octet of the $\pi$
and vector mesons of the octet of the $\rho$, and unitarizing in coupled
channels solving the coupled Bethe-Salpeter equations, one obtains the scattering
matrix for pseudoscalar mesons with vectors for different quantum numbers, 
which contains poles that can be associated to known resonances like the 
$a_1(1260)$, $b_1(1235)$, etc. The SU(3) decomposition of $8\times 8$ 
\begin{equation}
8 \times 8 = 1 + 8_s + 8_a + 10 + \overline{10} + 27 
\end{equation}
leads here to two octets, unlike in the case of the interaction of pseudoscalars
among themselves where there is only room for the $8_s$ representation. This is
why here one finds different G-parity states like the $a_1$ and $b_1$,
 the $f_1(1285)$, $h_1(1380)$, plus an
extra $h_1(1170)$ that one can identify with the singlet state. One should then
find two $K_1$ states, which do not have defined G-parity. One might
think that these states are the $K_1(1270)$ and the $K_1(1400)$ states. However
the theory fails to predict a state with such a large mass as the $K_1(1400)$
and with its decay properties. Instead, in \cite{roca} two states were found with 
masses close by,  given, after some fine tunning, by 1197 MeV and 1284 MeV, and 
widths of about 240 MeV and 140 MeV, respectively.  The interesting thing about
these states is that the first one couples most strongly to  $K^* \pi$, while 
the second state couples most strongly to $K \rho$. One could hope that these
two states could be observed experimentally. Indeed, this is the case as was
shown in the recent work \cite{geng} by looking at two reactions which have
either $K^* \pi$ or $K \rho$ in the final state and which clearly show the peak
at different positions, as one can observe in fig. \ref{fig:CERN3}.

 It is interesting to recall that in the experimental analysis done in
 \cite{daum} only one $K_1(1270)$ resonance was included (together with the
 $K_1(1400)$ which shows up at higher energies), but the background was very
 large and the peaks appeared from interference of large background terms rather
 than from the effect of the resonance.   Instead in \cite{geng}, with the
 introduction of the two resonances obtained in our approach and the 
 background generated by the same chiral unitary approach, together with 
  a contribution from
 the $K_1(1400)$ considered phenomenologically, the description of all data in
 fig \ref{fig:CERN3} follows in a natural way. 

\section{Dynamically generated scalar mesons with open and hidden charm  }

 A generalization to SU(4) of the SU(3) chiral Lagrangian for meson-meson 
 interaction is done in \cite{daniel} to study meson-meson  interaction
 including charm. The breaking of SU(4) is done as in \cite{hofmannbar,mizutani},
 where the crossed exchange of vector mesons is employed as it accounts 
 phenomenologically  for the Weinberg-Tomozawa term in the chiral Lagrangians. 
  With this in mind, when the exchange is due to a heavy vector
 meson the corresponding term is corrected by the ratio of square masses of the
 light vector meson to the heavy one (vectors with a charmed quark).  We also
 use a different pattern of SU(4) symmetry breaking by following the lines  of
 a chiral motivated model with general SU(N) breaking \cite{walliser}.  The picture
 generalizes the model used in \cite{koloscalar,hofmann,chiang}, where only the light vector
 mesons are exchanged. The same states generated in \cite{lutz,chiang} are  also
 generated in \cite{daniel} with some changes, but in addition one obtains
 states with hidden charm. The changes refer to the states of the sextet, which
 in  \cite{daniel} appear rather broad, while in the other works are narrow states. In
 table 1 we show the states with charm or hidden charm
 obtained in the approach. 

\begin{table}
\begin{center}
Table 1: {Pole positions for the model. The column Irrep shows the 
results in the $SU(3)$ limit.} \label{poleposset2}\\
\begin{tabular}{c|c||c|c|c|c|c}
\hline
C& Irrep &S&I$(J^{P})$& RE($\sqrt{s}$) (MeV)& IM($\sqrt{s}$) (MeV)&Resonance ID\\
 & Mass (MeV) & & & & & \\
\hline
\hline
 &          & & & & & \\
1&$\bar 3$&1&0$(0^+)$&2317.25&0&$D_{s0}^*(2317)$ \\
\cline{3-7}
 &2327.96 &0&$\oh(0^+)$&2129.26&-157.00&$D_0^*(2400)$ \\
\cline{2-7}
 & 6   &1&1$(0^+)$&2704.31&-459.50& (?) \\
\cline{3-7}
& 2394.87&0&$\oh(0^+)$&2694.69&-441.89& (?) \\
\cline{3-7}
& -i219.33&-1&0$(0^+)$&2709.39&-445.73&(?) \\
\hline
0&1&0&0$(0^{+})$&3718.93&-0.06&(?)\\
& & & & & & \\
\hline
\end{tabular}
\end{center}
\end{table}
As we can see, the $D_{s0}(2317)$ and $D_0(2400)$ appear in the approach, the
last one at lower energies than experiment, but consistent with the data
considering the large width of the state and the theoretical and experimental
uncertainties on the mass. The other three charm states in the table come from a
sextet and they are very broad in our approach ($\Gamma \sim IM(\sqrt{s})$). 

  The very interesting and novel thing with respect to other theoretical works
 is the heavy state with zero charm . It is a hidden charm state mostly built from
 $D \bar{D}$ and $D_s \bar{D}_s$. The fact that this state has such a narrow
 width in spite of having all the meson-meson states of the light sector open
 for decay, is an interesting consequence of the work, which largely decouples
 the light sector from the heavy one respecting basic OZI rules.  There is no
 experimental information on this state now, but an
 enhancement of the cross section of the $e^+ e^- \to J/\psi D \bar{D}$ close to
 threshold seen in \cite{plakhov} could be interpreted in \cite{danielnew} 
 as a consequence of the
 effect close to  $D \bar{D}$ threshold of the X(3700), a bound state below
 threshold.

\section{Dynamically generated axial vector mesons with open and hidden charm}

With the interaction of pseudoscalar mesons with vector mesons in
\cite{danielaxial} one obtains the results shown in Table 2. In addition to
the well known $D_{s1}(2460)$, $D_1(2430)$, $D_{s1}(2536)$ and $D_1(2420)$ (and
all those in the light sector already found in \cite{lutzaxial,roca}) one obtains
new states, which could be observed, although some of them are either too broad
or correspond to cusps. 

\begin{table}
Table 2: Pole positions for the model. The column Irrep shows the results in 
the $SU(3)$ limit. The results in brackets for the $Im\sqrt{s}$ are obtained 
taking into account the finite width of the $\rho$ and $K^*$ mesons. \label{axialsmodel}
\begin{center}
\begin{tabular}{c|c||c|c|c|c|c}
\hline
C& Irrep &S&I$^G(J^{PC})$& RE($\sqrt{s}$) & IM($\sqrt{s}$) &Resonance ID\\
 & Mass (MeV) & & & (MeV)& (MeV)& \\
\hline
\hline
1&$\bar 3$&1&0$(1^+)$&2455.91&0&$D_{s1}(2460)$ \\
\cline{3-7}
 &2432.63 &0&$\oh(1^+)$&2311.24&-115.68&$D_1(2430)$ \\
\cline{2-7}
 & 6   &1&1$(1^+)$&2529.30&-238.56& (?) \\
\cline{3-7}
& 2532.57&0&$\oh(1^+)$&Cusp (2607)&Broad& (?) \\
\cline{3-7}
& -i199.36&-1&0$(1^+)$&Cusp (2503)&Broad&(?) \\
\cline{2-7}
& $\bar 3$ &1&0$(1^+)$&2573.62&-0.07~[-0.07]&$D_{s1}(2536)$ \\
\cline{3-7}
&2535.07&0&$\oh(1^+)$&2526.47&-0.08~[-13]&$D_1(2420)$ \\
& -i0.08& &     &      &     & \\
\cline{2-7}
& 6 & 1&1$(1^+)$& 2756.52&-32.95~[cusp]&(?) \\
\cline{3-7}
&Cusp (2700)&0&$\oh(1^+)$&2750.22&-99.91~[-101]&(?) \\
\cline{3-7}
&Narrow&-1&0$(1^+)$&2756.08&-2.15~[-92]&(?) \\
\hline
0&1&0&0$^+(1^{++})$&3837.57&-0.00&$X(3872)$\\
&3867.59& & & & & \\
\cline{2-7}
&1&0&0$^-(1^{+-})$&3840.69&-1.60&(?)\\
&3864.62& & & & & \\
\hline
\end{tabular}
\end{center}
\end{table}

     Very interesting and novel of the present approach is the generation of the
X(3872) with positive C-parity and another state nearly degenerate
with negative C-parity. It would be interesting to see if a state with negative
C-parity is observed, but the large branching fraction 
\begin{equation}
\frac{B(X \to \pi^+ \pi^- \pi^0 J/\psi)}{B(X \to \pi^+ \pi^-  J/\psi)}=1.0\pm
0.4 \pm 0.3
\end{equation}
indicates either   a very large G-parity (isospin) violation (quite unlikely), or the existence of
another state with different C-parity (G-parity also in this case).

\section{Dynamically generated $1/2^+$ baryon states from the interaction of
two mesons and one baryon}

We discussed before how the low lying $1/2^-$ baryon resonances appear
dynamically generated in the chiral unitary approach. The low lying $1/2^+$
resonances are not less problematic and quark models have difficulties in
reproducing them \cite{glozman}. Experimentally some of them are poorly
understood, few of them 
 possess four-star status and three possess three-star status.
Among the rest some resonances are listed with unknown spin parity and two
are controversial in nature. The situation is slightly better with the
$\Lambda$ resonances in the same energy region, except for the $\Lambda$(1600)
and $\Lambda$(1810), where the peak positions and widths, obtained by different
partial wave analysis groups, vary a lot. Many of these S=$-1$ states seem to
have significant branching ratios for three-body, i.e., two meson-one baryon,
decay channels. However, no theoretical attempt has been made to study the
three body structure of these resonances, until recently when a coupled channel
calculation for two meson one baryon system was carried out using chiral
dynamics \cite{mko}. 

\section{Formalism for the three body systems} 

We take advantage of the fact that there are strong correlations in the meson
baryon sector in L=0, and with S=$-1$ one obtains many $1/2^-$ resonances.
The $\Lambda(1405)\,S_{01}$ ($J^P=1/2^-$) couples strongly
to the $\pi-\Sigma$ and its coupled channels. Considering this we build the
three body coupled channels by adding a pion to combinations of a pseudoscalar
meson of the $0^-$  SU(3) octet and a baryon of the $1/2^+$ octet which couple
to $S=-1$. For the total charge zero of the three body system we get twenty-two
coupled channels.

To solve the Faddeev equations we write the two body $t$-matrices using unitary
chiral dynamics. These $t$-matrices can be split into an on-shell part,
depending only on the respective center of mass energy,  and an off-shell part,
which is inversely proportional to the propagator of the off-shell particle.
This off-shell part cancels a propagator in the three body scattering diagrams,
leading to a diagram with a topological structure equivalent to that of a three
body force \cite{mko}. To this, one must add the three body forces originating
directly from the chiral Lagrangians. Interestingly, in our case, we find the
three forces from the two sources to get canceled in the SU(3) limit and in
case of low momentum transfer to the baryon. In a realistic case, we find them
to sum-up to merely 5 \% of the total on-shell contribution of the
$t$-matrices to the Faddeev equations. The formalism is thus developed further
in terms of the on-shell parts of the two body $t$-matrices.

We begin with Faddeev equations 
\begin{equation}
T^i=t^i\delta^3(\vec{k}^{\,\prime}_i-\vec{k}_i)+ t^i g [ T^j + T^k ],
\end{equation}
which if iterated while neglecting the terms with $\delta^3(\vec{k}^{\,\prime}_i-\vec{k}_i)$,
which correspond to the disconnected diagrams, will give
\begin{eqnarray}\nonumber 
T^i = t^i g^{ij} t^j + t^i g^{ik} t^k + t^i g^{ij} t^j g^{jk} t^k + t^i g^{ij} t^j g^{ji} t^i
+t^i g^{ik} t^k g^{kj} t^j + t^i g^{ik} t^k g^{ki} t^i + ... .
\end{eqnarray}

In order to factorize the Faddeev equations one writes the terms with three
successive interactions explicitly, which already involve a loop evaluation. One
finds technically how to go from the diagrams with two interactions to
those with three interactions and the algorithm found is then used for the next
iterations, leading thus to a set of algebraic equations, which are solved
within twenty two coupled channels.

\begin{figure}[hbt]
\includegraphics[scale=0.68]{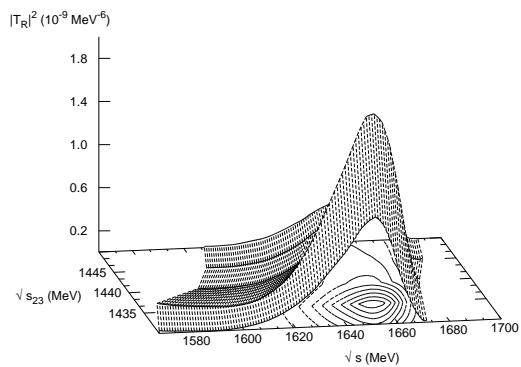}
\caption[]{The $\Sigma$ (1660) resonance in the $\pi^0 \pi^0 \Sigma^0$ channel.}\label{fig1}
\end{figure}

The resulting equations have been solved with the input two body $t$-matrices
obtained by solving the Bethe-Salpeter equation as in 
\cite{npa,angels,inoue}. We find four $\Sigma$ and two $\Lambda$ states
 \cite{pdg} as
dynamically generated resonances in the two meson-one baryon systems, implying
a strong coupling of the S=$-1$ resonances, in this region, to the three body
decay channels. In Fig. \ref{fig1}, we show one of the resonances, corresponding to the
$\Sigma$(1660) \cite{pdg} found in our study in the squared amplitude for the
$\pi^0 \pi^0 \Sigma^0$ channel. In addition to this, we find evidence
for (1) another $1/2^+$ resonance, i.e., the $\Sigma$(1770), (2) for the
controversial $\Sigma$(1620) and (3) for the $\Sigma$(1560), which is listed
with unknown spin-parity \cite{pdg}. In the isospin 0 sector we find evidence
for the $\Lambda$ (1600) and $\Lambda$ (1810). To conclude, we are finding a
new picture for the low lying $1/2^+$ baryon states which largely correspond to
bound states or resonances of two mesons and a baryon.

\section*{Acknowledgments}

L. S. Geng wishes to acknowledge support from the Ministerio de Educacion in the
program of Doctores y Tecnologos extranjeros and D. Strottman in the one of
sabbatical. A. Martinez and D. Gamermann from the Ministry of
Education and Science and K. Khemchandani from the HADRONTH project for the EU.
 This work is partly supported by
DGICYT Contract No. BFM2003-00856, FPA2007-62777 , the Generalitat Valenciana, and the
E.U. FLAVIAnet network Contract No. HPRN-CT-2002-00311. This
research is part of the EU Integrated Infrastructure Initiative
Hadron Physics Project under Contract No. RII3-CT-2004-506078.

\end{document}